\documentclass[twocolumn]{aastex701}

\usepackage{amsmath,amssymb}
\usepackage{xspace}
\usepackage{CJK}

\usepackage{hyperref}
\hypersetup{
    colorlinks=true,
    linkcolor=blue,
    filecolor=magenta,      
    urlcolor=cyan,
}

\usepackage{color}

\newcommand{\um}{\ensuremath{\mu\rm{m}}\xspace}

\newcommand{\alphaco}{\ensuremath{\alpha_{\rm{CO}}}\xspace}
\newcommand{\gdr}{\ensuremath{\delta_{\rm{GDR}}}\xspace}

\newcommand{\uJy}{\ensuremath{\mu\rm{Jy}}\xspace}

\newcommand{\Tdust}{\ensuremath{\rm{T}_{\rm{dust}}}\xspace}
\newcommand{\Mdust}{\ensuremath{M_{\rm{dust}}}\xspace}

\newcommand{\Mstar}{\ensuremath{M_{\rm{star}}}\xspace}

\newcommand{\MHt}{\ensuremath{M_{\rm{H}_2}}\xspace}

\newcommand{\Msun}{\ensuremath{\rm{M}_\odot}\xspace}

\newcommand{\fht}{\ensuremath{f_{\rm{H}_2}}\xspace}
\newcommand{\fdust}{\ensuremath{f_{\rm{dust}}}\xspace}
\newcommand{\tdep}{\ensuremath{t_{\rm{dep}}}\xspace}

\newcommand{\percc}{\ensuremath{\rm{cm}^{-3}}\xspace}

\newcommand{\Tex}{\ensuremath{T_\mathrm{ex}}\xspace}

\newcommand{\ci}{[C{\scriptsize I}]\xspace}
\newcommand{\cii}{[C{\scriptsize II}]\xspace}

\newcommand{\oii}{[O{\scriptsize II}]\xspace}

\newcommand{\arc}{\ensuremath{''}\xspace}

\shortauthors{J.~S.~Spilker, et~al.}
\shorttitle{Weird Gas to Dust Ratios in High-z Quiescent Galaxies}

\begin{document}
\begin{CJK*}{UTF8}{gbsn}

\defcitealias{spilker18b}{JS18}
\defcitealias{whitaker21b}{KW21b}

\title{Unusually High Gas-to-Dust Ratios Observed in High-Redshift Quiescent Galaxies}

\correspondingauthor{Justin S. Spilker}
\email{jspilker@tamu.edu}

\author[0000-0003-3256-5615]{Justin~S.~Spilker}
\affiliation{Department of Physics and Astronomy and George P. and Cynthia Woods Mitchell Institute for Fundamental Physics and Astronomy, Texas A\&M University, 4242 TAMU, College Station, TX 77843-4242, US}
\email{jspilker@tamu.edu}

\author[0000-0001-7160-3632]{Katherine E. Whitaker}
\affiliation{Department of Astronomy, University of Massachusetts, Amherst, MA 01003, USA}
\affiliation{Cosmic Dawn Center at the Niels Bohr Institute, University of Copenhagen\
 and DTU-Space, Technical University of Denmark}
\email{kwhitaker@astro.umass.edu}

\author[0000-0002-7064-4309]{Desika~Narayanan}
\affiliation{Department of Astronomy, University of Florida, 211 Bryant Space Science Center, Gainesville, FL 32611 USA}
\affiliation{Cosmic Dawn Center at the Niels Bohr Institute, University of Copenhagen\
 and DTU-Space, Technical University of Denmark}
\email{desika.narayanan@ufl.edu}

\author[0000-0001-5063-8254]{Rachel~Bezanson}
\affiliation{Department of Physics and Astronomy and PITT PACC, University of Pittsburgh, Pittsburgh, PA 15260, USA}
\email{rachel.bezanson@pitt.edu}

\author[0000-0001-8246-1676]{Sarah~Bodansky}
\affiliation{Department of Astronomy, University of Massachusetts, Amherst, MA 01003, USA}
\email{sbodansky@umass.edu}

\author[0000-0002-1759-6205]{Vincenzo~R.~D'Onofrio}
\affiliation{Department of Physics and Astronomy and George P. and Cynthia Woods Mitchell Institute for Fundamental Physics and Astronomy, Texas A\&M University, 4242 TAMU, College Station, TX 77843-4242, US}
\email{donofr19@tamu.edu}

\author[0000-0002-1109-1919]{Robert~Feldmann}
\affiliation{Department of Astrophysics, Universit\"at Z\"urich, CH-8057, Zurich, Switzerland}
\email{feldmann@physik.uzh.ch}

\author[0000-0003-4700-663X]{Andy~D.~Goulding}
\affiliation{Department of Astrophysical Sciences, Princeton University, Princeton, NJ 08544, USA}
\email{goulding@astro.princeton.edu}

\author[0000-0002-5612-3427]{Jenny~E.~Greene}
\affiliation{Department of Astrophysical Sciences, Princeton University, Princeton, NJ 08544, USA}
\email{jgreene@astro.princeton.edu}

\author[0000-0002-7613-9872]{Mariska Kriek}
\affiliation{Leiden Observatory, Leiden University, P.O. Box 9513, 2300 RA Leiden, The Netherlands}
\email{kriek@strw.leidenuniv.nl}

\author[0000-0002-0696-6952]{Yuanze~Luo}
\affiliation{Department of Physics and Astronomy and George P. and Cynthia Woods Mitchell Institute for Fundamental Physics and Astronomy, Texas A\&M University, 4242 TAMU, College Station, TX 77843-4242, US}
\email{yluo37@tamu.edu}

\author[0000-0003-4075-7393]{David~J.~Setton}
\altaffiliation{Brinson Prize Fellow}
\affiliation{Department of Astrophysical Sciences, Princeton University, Princeton, NJ 08544, USA}
\email{davidsetton@princeton.edu}

\author[0000-0002-1714-1905]{Katherine~A.~Suess}
\affiliation{Department for Astrophysical \& Planetary Science, University of Colorado, Boulder, CO 80309, USA}
\email{suess@colorado.edu}

\author[0000-0002-5027-0135]{Arjen~van~der~Wel}
\affiliation{Sterrenkundig Observatorium, Universiteit Gent, Krijgslaan 281 S9, B-9000 Gent, Belgium}
\email{arjen.vanderwel@ugent.be}

\author[0000-0003-1535-4277]{Margaret E. Verrico}
\affiliation{University of Illinois Urbana-Champaign Department of Astronomy, University of Illinois, 1002 W. Green St., Urbana, IL 61801, USA}
\affiliation{Center for AstroPhysical Surveys, National Center for Supercomputing Applications, 1205 West Clark Street, Urbana, IL 61801, USA}
\email{verrico2@illinois.edu}

\author[0000-0003-2919-7495]{Christina~C.~Williams}
\affiliation{NSF's National Optical-Infrared Astronomy Research Laboratory, 950 N. Cherry Avenue, Tucson, AZ 85719, USA}
\affiliation{Steward Observatory, University of Arizona, 933 North Cherry Avenue, Tucson, AZ 85721, USA}
\email{christina.williams@noirlab.edu}

\author[0000-0001-5962-7260]{Charity~Woodrum}
\affiliation{Observational Cosmology Lab, Code 665, NASA Goddard Space Flight Center, 8800 Greenbelt Rd., Greenbelt, MD 20771, USA }
\email{charity.a.woodrum@nasa.gov}

\author[0000-0002-9665-0440]{Po-Feng~Wu}
\affiliation{Graduate Institute of Astrophysics, National Taiwan University,
Taipei 10617, Taiwan}
\email{wupofeng@phys.ntu.edu.tw}

%%%%%%%%%%%%%%%%%%%%%%%%%%%%%%%%%%%%%%%%%%%%%%%%%%%%%%%%%%%%%%%%%%%%%%%%%%%%%%%%%%%%%
%%%%%%%%%%%%%%%%%%%%%%%%%%%%%%%%%%%% ABSTRACT %%%%%%%%%%%%%%%%%%%%%%%%%%%%%%%%%%%%%%%
%%%%%%%%%%%%%%%%%%%%%%%%%%%%%%%%%%%%%%%%%%%%%%%%%%%%%%%%%%%%%%%%%%%%%%%%%%%%%%%%%%%%%
\begin{abstract}

Tracking the cold molecular gas contents of galaxies is critical to understand the interplay between star formation and galaxy growth across cosmic time. Observations of the long-wavelength dust continuum, a proxy for the cold gas, are widely used in the high-redshift community because of their ease and efficiency. These measurements rely on the assumption of a molecular gas-to-dust mass ratio, typically taken to be $\gdr \approx 100$ in massive, metal-rich systems. We present Atacama Large Millimeter/submillimeter Array (ALMA) observations of the 870\,\um dust continuum in a sample of five massive quiescent galaxies at $z\sim1$ with existing detections of CO(2--1). We find surprisingly weak dust emission, falling a factor of $\gtrsim$0.4--0.8\,dex below the typical correlation between CO and continuum luminosity. We interpret this dust deficiency as evidence for unusually high \gdr in these galaxies, which we calculate to range from 300 to at least 1200. Our results and other observations from the literature are generally compatible with predictions from the SIMBA cosmological simulation that dust is preferentially destroyed in quiescent galaxies. Ultimately, we conclude that the dust continuum is a highly unreliable tracer of the molecular gas in high-redshift quiescent galaxies. As a consequence we may know much less about the cold gas contents of this population than previously thought.

\end{abstract}

%%%%%%%%%%%%%%%%%%%%%%%%%%%%%%%%%%%%%%%%%%%%%%%%%%%%%%%%%%%%%%%%%%%%%%%%%%%%%%%%%%%%%
%%%%%%%%%%%%%%%%%%%%%%%%%%%%%%%%%% Introduction %%%%%%%%%%%%%%%%%%%%%%%%%%%%%%%%%%%%%
%%%%%%%%%%%%%%%%%%%%%%%%%%%%%%%%%%%%%%%%%%%%%%%%%%%%%%%%%%%%%%%%%%%%%%%%%%%%%%%%%%%%%
\section{Introduction} \label{intro}

Because new stars form in clouds of cold, molecular gas, the amount of this gas is one of the key properties that gives insight into the past and future evolution of galaxies. Following significant observational investments by radio, sub/millimeter, and far-infrared telescopes, we now have a decent understanding of how the amount of cold gas in typical massive star-forming galaxies changes with stellar mass (\Mstar), star formation rate (SFR) or specific SFR (sSFR$\,\equiv\,$SFR/\Mstar), and redshift \citep[for reviews, see e.g.][]{hodge20,tacconi20}. 

Molecular hydrogen itself lacks a permanent electric dipole moment and its quadrupole transitions are not significantly excited in the cold interstellar medium (ISM), so indirect tracers are needed to estimate the molecular mass \MHt. Millimeter-band spectral lines of CO have been used for decades for this purpose \citep[e.g.][]{bolatto13}, but there has also been recent interest in several atomic fine structure lines in particular at high redshifts \citep[e.g. \ci, \cii;][]{carilli13}. However, thanks to wide-field bolometer cameras and sensitive interferometers, often the quickest way to reach a given limiting gas mass is to observe the long-wavelength continuum emission from dust heated and intermixed with H$_2$ in the ISM \citep[e.g.][]{magdis12,scoville16,tacconi20}.

These methods rely on conversion factors to translate the observed line or continuum luminosity into \MHt. For CO observations, one must estimate the molecular excitation if the ground-state transition has not been measured, and adopt or estimate a CO to H$_2$ conversion factor \alphaco. Thanks to decades of study, the variations in \alphaco are reasonably well understood \citep[e.g.][]{bolatto13}; compared to the typical value in the Milky Way ($\approx$4.4\,\Msun/(K\,km/s\,pc$^2$)), \alphaco increases in low-metallicity galaxies and decreases in warm, gas-rich galaxies where the line opacity is lower. For dust continuum, the mass-weighted dust temperature \Tdust and emissivity of the dust grains determine the mass of the dust, and a gas-to-dust mass ratio $\gdr \equiv \MHt/\Mdust$ is used to translate this into \MHt.\footnote{Some methods wrap more than one of these factors into a single conversion, e.g. \citealt{scoville16}.} The value of \gdr is $\approx$100--200 in metal-rich galaxies with a scatter of $\approx$0.4\,dex, and in the nearby universe, values $\gdr \gtrsim 1000$ are only observed in low-metallicity (less than about half solar) systems \citep{leroy11,remyruyer14,devis19,galliano21,park24a}.

Because they are typically gas-poor and therefore faint in these observables, relatively little is known about the cold gas contents of quiescent galaxies beyond the local universe \citep[e.g.][]{young11,davis16}. Galaxies take diverse pathways to quiescence, and so their gas contents may be expected to be similarly diverse. Recently-quenched galaxies can host large molecular reservoirs that decline with time, occasionally in spatial distributions that reflect recent interactions \citep[e.g.][]{french15,alatalo16,bezanson22,spilker22b,wu23,zanella23,donofrio25}. Older quiescent galaxies also appear to have a wide variety of cold gas properties. Some individual sources are found to be very gas-poor \citep[e.g.][]{sargent15,bezanson19,williams21}, others relatively gas-rich \citep[e.g.][]{rudnick17,hayashi18,lee24,pigarelli25,umehata25}, while other studies find a mixture or wide range (e.g. \citealt{spilker18b}, hereafter \citetalias{spilker18b}). It is interesting to note that many of the gas-rich quiescent galaxies are found in overdensities, but the role of galaxy environment is also not yet well understood.

Observing either CO or dust continuum in gas-poor galaxies can be costly, so there is not yet a large, homogeneously-selected sample of high-redshift quiescent galaxies with molecular gas constraints. Stacking of large samples of far-IR and submillimeter images of quiescent galaxies can probe the typical gas contents, and early studies found modestly large gas fractions $\fht \equiv \MHt/\Mstar \approx 0.05-0.1$ \citep{gobat18,magdis21}. The datasets used had poor spatial resolution, but subsequent stacking experiments have found similar \citep{blanquezsese23} to a factor of a few lower gas fractions \citep{adscheid25}. The ISM-rich stacking results are in mild tension with some studies of individual galaxies at similar redshifts (e.g. \citealt{whitaker21a,williams21}), which find much lower $\fht \lesssim 0.01$. Some of these differences are due to differing assumptions and modeling approaches, but the early low-resolution stacking results still recover higher gas fractions by a factor of a few even when the same framework is applied to all measurements.

As we look to improve our knowledge of molecular gas in quiescent galaxies throughout the universe, we must understand whether the relevant conversion factors still hold true. In this context recent results from the SIMBA cosmological hydrodynamical simulation are concerning \citep{dave19,li19}. While SIMBA, which separately tracks H$_2$ and dust, can produce some quiescent galaxies with large dust masses as a consequence of prolonged grain growth in the ISM \citep{lorenzon25a}, they are more commonly very dust-poor. Moreover, the value of \gdr can increase far above the typically-assumed value to $\gdr > $100,000, with orders-of-magnitude scatter (\citealt{whitaker21b}, hereafter \citetalias{whitaker21b}). Both of these effects are mostly due to thermal sputtering of dust grains by hot electrons, which destroys dust grains, and a subsequent lack of dust re-growth in quiescent galaxies. If confirmed, this would imply that detections (or non-detections) of the far-IR dust continuum say virtually nothing about the cold gas contents of quiescent galaxies.

In this paper, we present new ALMA 870\,\um follow-up observations of five $z\sim1$ quiescent galaxies that had previous detections of CO(2--1) in order to constrain \gdr and test the predictions from SIMBA. Section~\ref{data} describes the observations and analysis methods we use. In Section~\ref{results} we show that these galaxies indeed have unexpectedly weak dust emission, evidence that \gdr can reach more than 1000 in presumably metal-rich quiescent galaxies. We briefly conclude in Section~\ref{conclude}. We use a flat $\Lambda$CDM cosmology with $\Omega_m=0.3$ and $H_0=68$\,km/s\,Mpc$^{-1}$, as in SIMBA, and a \citet{chabrier03} initial mass function.

%%%%%%%%%%%%%%%%%%%%%%%%%%%%%%%%%%%%%%%%%%%%%%%%%%%%%%%%%%%%%%%%%%%%%%%%%%%%%%%%%%%%%
%%%%%%%%%%%%%%%%%%%%%%%%%%%%%%%% Data & Analysis %%%%%%%%%%%%%%%%%%%%%%%%%%%%%%%%%%%%
%%%%%%%%%%%%%%%%%%%%%%%%%%%%%%%%%%%%%%%%%%%%%%%%%%%%%%%%%%%%%%%%%%%%%%%%%%%%%%%%%%%%%
\section{Sample, Observations, and Simulations} \label{data}

%%%%%%%%%%%%%%%%%%%%%%%%%%%%%%%%%%%%%%%%%%%%%%%%%%%%%%%%%%%%%%%%%%%%%%%%%%%%%%%%%
%%% FIGURE - ALMA images
\begin{figure*}[t]
\begin{centering}
\includegraphics[width=\textwidth]{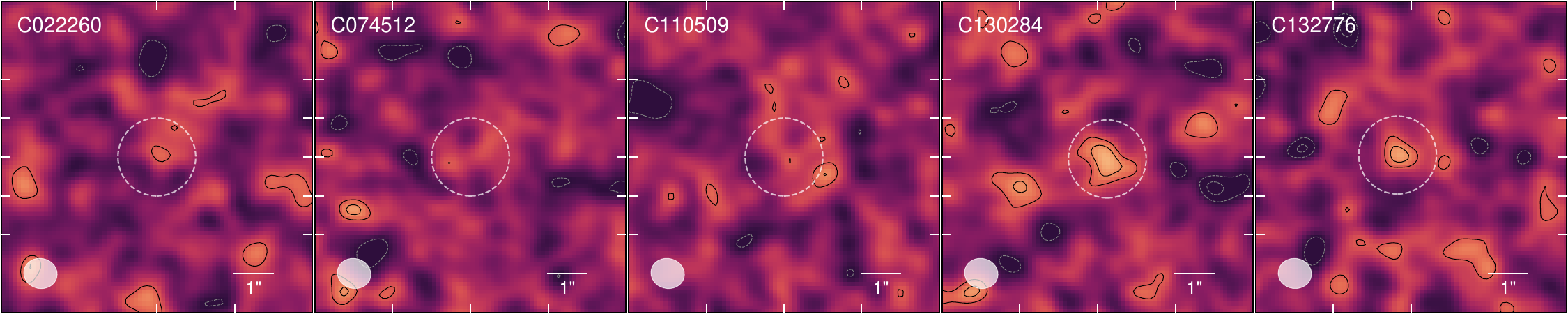}
\end{centering}
\caption{
Tapered ALMA 345\,GHz cutout images of our quiescent galaxy sample. Contours are drawn at $\pm$2, 3$\sigma$, where $\sigma$ varies from 20.6--22.6\uJy/beam. Only C130284 and C132776 are detected at $>3\sigma$. Dashed white circles indicate the 2'' photometric apertures.
}\label{fig:images}
\end{figure*}

\subsection{Quiescent Galaxies and ALMA Observations} \label{sample}

We selected a sample of five quiescent galaxies at $z\sim1$ with existing detections of CO(2--1) from the literature. Because the prediction from SIMBA \citepalias{whitaker21b} is that gas and dust are nearly uncorrelated at low sSFR, requiring a detection of CO does not impose a strong bias on \gdr. The sample consists of 4 galaxies at $z\sim0.7$ from the LEGA-C survey presented in \citetalias{spilker18b} and one object at $z=1.24$ presented in \citet{williams21}. More details on the parent samples, CO(2--1) measurements, and galaxy properties can be found in those works. In particular, we note that the LEGA-C objects have higher star formation rates (SFRs) than most other LEGA-C galaxies of comparable mass \citepalias{spilker18b}, which may be tied to low-level rejuvenation events that formed $\approx$0.3--6\% of the stellar mass \citep{woodrum22}. The typical mass-weighted stellar ages are 3--5\,Gyr \citep{williams21,woodrum22}. No metallicity estimates are available for these galaxies, but they are expected to be super-solar based on the mass--metallicity relation \citep{mannucci10}. Although we chose CO-detected galaxies for ALMA followup, the implied gas masses agree with extrapolations of literature scaling relations \citepalias{spilker18b}. No galaxy shows evidence of AGN from either the optical spectra or mid-IR colors \citep{woodrum22}, but one target (C110509) appears to have a radio AGN \citep{barisic17}.

Estimating SFRs at low levels is notoriously difficult. Several SFR estimates are available for our targets in the literature, based on the UV$+$IR luminosity, spectrophotometric fitting, and/or \oii luminosity. The spectrophotometric fitting results \citep{williams21,woodrum22} both use the same code and flexible star formation histories to estimate the recent SFR, but the modeling frameworks are not necessarily identical. For ease and consistency, we adopt the SFRs based on the UV$+$IR luminosity for both our own targets and literature comparison galaxies described below. In practice the ``IR'' portion of the UV$+$IR SFRs comes solely from 24\,\um imaging; neither far-IR nor ALMA observations are used. The UV$+$IR SFRs are typically comparable to or higher than other estimates, resulting in the `least quiescent' estimates of sSFR. We expect these to overestimate the true SFR because old stars can also contribute to the dust heating at low SFR \citep[e.g.][]{fumagalli14,utomo14,hayward15,leja19,wu23}. A summary of the galaxy properties is given in Table~\ref{tab:sample}.

ALMA observations of the rest-frame $\approx$500\,\um dust continuum were carried out in project 2022.1.00642.S (PI Spilker) from 2022 October 7--11. All five targets are located in the COSMOS field, and were observed for 10\,min on-source in each of four execution blocks with shared bandpass, amplitude, and complex gain calibrations. The correlator was configured to use the standard observatory Band~7 continuum setup centered at $\approx$345\,GHz. The array was in a compact configuration with baselines of 15--500\,m. We imaged the data using natural weighting of the visibilities, yielding a synthesized beam of $\approx$0.6\arc. We also applied a 0.5\arc uv-plane taper to produce images with $\approx$0.9\arc resolution, which we use in the remainder of this work. The typical sensitivity of the tapered images is $\approx$21\,\uJy/beam. We measure flux densities from 2.0\arc diameter apertures placed at the peak of the detected emission (C130284, C132776) or the phase center for undetected galaxies (C022260, C074512, C110509), and measure uncertainties using randomly-placed apertures of the same size. In all cases, the aperture centers coincide with the peak of the CO(2--1) emission. We tested both more aggressive uv tapers (up to 1.5\arc, yielding a $\approx$2.1\arc synthesized beam) and larger aperture sizes, and find consistent photometric measurements. ALMA images of the targets are shown in Figure~\ref{fig:images}.

% MORE DETAILS, IF USED
For the three individually-undetected sources, we stacked the images to reach modestly deeper limits. We opt for a simple image-based median stack because the depth and resolution of the input images are essentially the same. Before stacking, we rescale each image by a factor that accounts for the different redshifts of the galaxies \citep[e.g.][]{spilker14,scoville16}. This factor is $<$15\%,  because the cosmological dimming due to increasing source distance is nearly canceled by the steeply-rising dust SED (the well-known strongly negative `k-correction' in the submm). No continuum emission is detected in the stack, with a 1$\sigma$ sensitivity of 14\,\uJy/beam.

\subsection{Literature Comparison Galaxies}

We assembled a sample of $0.5<z<4$ quiescent galaxies with gas and/or dust measurements from the literature. We began from the sample of $z>1$ sources compiled by \citet{lee24}. We supplemented this sample with the remainder of the LEGA-C sources from \citetalias{spilker18b} (i.e. the CO-undetected sources from that work) and the objects from \citet{siegel25,umehata25}. We do not include sources selected to be post-starburst, recently-quenched, etc., because their ISM contents are known to change rapidly with post-burst age \citep{french15,bezanson22}. Finally, we also include the (dust-only) stacking results from \citet{gobat18,magdis21,blanquezsese23}.\footnote{We explicitly choose not to include the $z\sim0.3$ sample of \citet{donevski23}, because those quiescent galaxies were selected to have far-IR detections, i.e. were intentionally selected to be dust-rich.}

The stellar masses of the comparison galaxies, including the stacking results, range from $\log(\Mstar/\Msun) = 10.2-11.6$ with a median of 11.1. The selection criteria can vary among different studies, but typically include either sSFR or UVJ color-color cuts (or both). It is beyond the scope of this work to scrutinize these differences in detail. In keeping with the spirit of the selection of our CO followup targets, we retained only galaxies with sSFR more than 3$\times$ below the star-forming sequence at the redshift of each galaxy. We also excluded any galaxies for which the continuum observation probed rest-frame wavelength $<$200\,\um, intended to ensure that the dust mass is estimated from the Rayleigh-Jeans tail of the dust SED.  We self-consistently estimate the gas and dust masses using the methods summarized in Section~\ref{gasdustmethod}.

%%%%%%%%%%%%%%%%%%%%%%%%%%%%%%%%%%%%%%%%%%%%%%%%%%%%%%%%%%%%%%%%%%%%%%%%%%%%%%%%%
%%% TABLE - sample properties
% Name  z   logMstar    SFR    S870    Mdust     MH2    GDR
\begin{deluxetable*}{lcccccCCCC}
\tablecaption{Summary of quiescent galaxy sample properties \label{tab:sample}}
\tablehead{
\colhead{Source}               & 
\colhead{RA}                   &
\colhead{dec}                  &
\colhead{z}                    &
\colhead{$\log(\Mstar/\Msun)$} &
\colhead{SFR$_{\mathrm{UV+IR}}$} &
\colhead{$S_{870\um}$}         &
\colhead{\Mdust}               &
\colhead{\MHt}                 &
\colhead{\gdr}                 \\
\colhead{}                     &
\colhead{}                     &
\colhead{}                     &
\colhead{}                     &
\colhead{}                     &
\colhead{\Msun/yr}             &
\colhead{\uJy}                 &
\colhead{10$^7$\,\Msun}        &
\colhead{10$^9$\,\Msun}        &
\colhead{}
}
\startdata
C022260 & 09:59:16.38 & +02:33:41.8 & 1.240 & 11.50 & 3.6 & <44      & <1.3      & 10.6\pm2.2 & >780      \\
C074512 & 10:01:42.88 & +02:01:21.9 & 0.733 & 11.15 & 6.3 & <42      & <1.1      &  6.6\pm1.7 & >610      \\
C110509 & 10:01:04.44 & +02:04:37.2 & 0.667 & 11.34 & 6.5 & <42      & <1.0      &  8.2\pm1.3 & >830      \\
C130284 & 10:00:13.78 & +02:19:37.0 & 0.602 & 11.18 & 6.8 & 150\pm31 & 3.3\pm0.7 & 10.0\pm1.3 & 300\pm70  \\
C132776 & 10:00:12.43 & +02:21:21.9 & 0.750 & 11.15 & 7.9 &  80\pm23 & 2.0\pm0.6 & 14.3\pm3.1 & 700\pm250 \\
stack   & ---         & ---         & 0.717 & 11.34 & 5.5 & <28      & <0.7      &  8.2\pm1.6 & >1200     \\
\enddata
\tablecomments{Redshifts, stellar masses, SFRs, and \MHt are from \citet{williams21} for C022260, and \citet{spilker18b} otherwise. The stack of individually undetected sources (`stack') uses the weighted average for the physical properties. All limits are 2$\sigma$.}
\end{deluxetable*}

\subsection{SIMBA Simulation Products} \label{simba}

We compare our observational results to the SIMBA cosmological hydrodynamical simulation, which includes a model for dust production, growth, and destruction physics. Full details on SIMBA and its dust modeling are given in \citet{dave19}, \citet{li19}, and \citet{narayanan21}. Briefly, molecular gas is computed with the \citet{krumholz11} prescription in dense $n_H > 0.13$\,\percc gas. The dust content is governed by the yields produced by Type~II supernovae and AGB stars, grain growth in the ISM, and destruction by thermal sputtering, astration, and supernova shocks (sputtering is the dominant destruction channel; \citealt{li19}, \citetalias{whitaker21b}). SIMBA's treatment of dust reproduces observational trends, including trends of \gdr with metallicity in star-forming systems to $z=2$ \citep{shapley20} and the dust mass function at $z=0-2$ \citep{li19,dudzeviciute20}.

In this work, we use the same simulation output as \citetalias{whitaker21b} and replicate their analysis for consistency. From the $z=1$ snapshot of the (100 $h^{-1}$ Mpc)$^3$ simulation box, we select massive galaxies with $\log(\Mstar/\Msun) > 10$. We remove any galaxies with no dust, H$_2$, or metals. The SIMBA comparison sample comprises 2358 galaxies with $\log(\mathrm{sSFR}/\mathrm{yr}^{-1}) > -10$ and 347 below this value. As in the observations, we assume that dust traces only molecular H$_2$ gas, ignoring neutral atomic gas. Because of known offsets between observed and simulated SFRs \citep{akins22}, we adjust the SIMBA SFRs to match the star-forming sequence of \citet{whitaker14}, as in \citetalias{whitaker21b}. At the high stellar masses $\log(\Mstar/\Msun) > 10$ we consider here, this adjustment is small, a $\approx$0.1--0.2\,dex upward revision to the SIMBA SFRs.

\subsection{Dust and Gas Mass Calculations} \label{gasdustmethod}

We recompute dust and gas masses for our sample and the literature comparison galaxies and stacks. We estimate dust masses following standard practice in the literature, namely $\Mdust = \frac{S_{\nu_{\rm{obs}}} D_L^2}{\kappa_{\nu_{\rm{rest}}} (1+z)} (B_{\nu_{\rm{rest}}}(\Tdust) - 
B_{\nu_{\rm{rest}}}(T_{\mathrm{CMB}}))^{-1}$, where $B_\nu(T)$ is the Planck function evaluated at temperature $T$ and $\kappa_\nu \equiv \kappa_{\nu_0} (\nu/\nu_0)^\beta$ is the dust absorption coefficient. We adopt the emissivity of \citet{dunne03}, with $\kappa_{\nu_0} = 2.64$\,m$^2$/kg at $\nu_0 = 2.4$\,THz varying with frequency with $\beta = 2.0$. We assume a mass-weighted $\Tdust = 25$\,K and note that masses vary approximately as $1/\Tdust$; see \citet{scoville16} for an extensive discussion of this point.

All of our target galaxies used CO(2--1) detections to constrain the molecular gas masses, but literature studies have used other CO transitions or \ci as alternate tracers. For the CO studies (including our own targets), we assume CO line excitation ratios $r_{21} = 0.8$, $r_{31} = 0.5$, and a Milky Way-like CO--H$_2$ conversion factor $\alphaco = 4.4$\,\Msun/(K\,km/s\,pc$^2$). Gas masses would increase by assuming either a higher \alphaco value or lower CO line ratios. For our CO(2--1) observations, the uncertainty in CO excitation is far subdominant to that in \alphaco. For the \ci studies, we adopt a \ci excitation temperature $\Tex = 25$\,K and a \ci/H$_2$ abundance ratio of $3\times10^{-5}$, following \citet{weiss05}. Molecular gas masses would increase if the \Tex is lower or if \ci is less abundant relative to H$_2$, with the carbon abundance as the main uncertainty for any reasonable value of \Tex. We discuss the consequences of these assumptions in Section~\ref{caveats}.

\section{Results} \label{results}

\subsection{A Deficit of Dust Emission}

%%%%%%%%%%%%%%%%%%%%%%%%%%%%%%%%%%%%%%%%%%%%%%%%%%%%%%%%%%%%%%%%%%%%%%%%%%%%%%%%%
%%% FIGURE - LCO vs Ldust
\begin{figure}[t]
\begin{centering}
\includegraphics[width=\columnwidth]{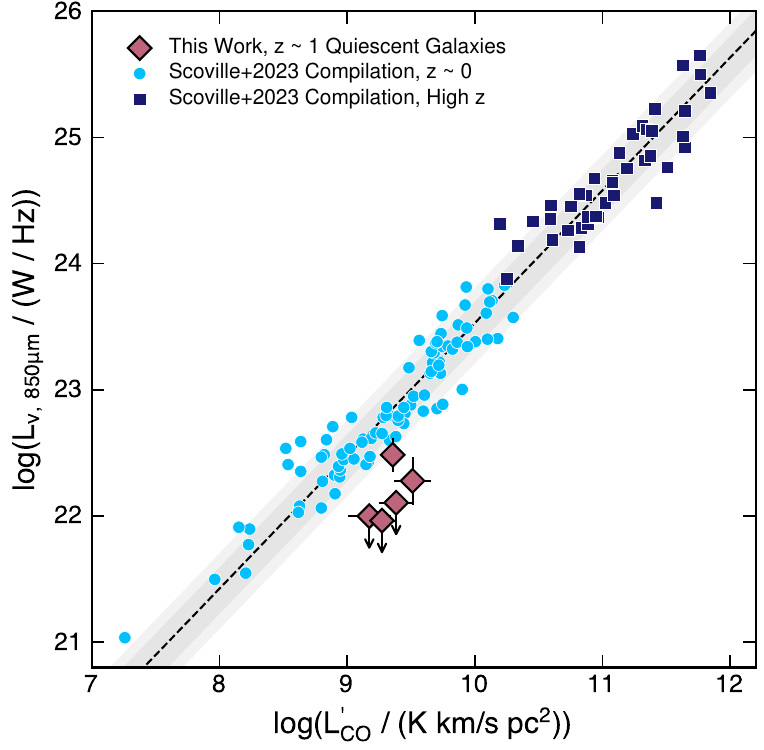}
\end{centering}
\caption{
Our sample of $z\sim1$ quiescent galaxies (diamonds) is empirically faint in dust continuum emission, regardless of any calibration or conversion factors. Compared to star-forming galaxies at low (circles) or high redshifts (squares), our sample galaxies have rest-frame 850\,\um continuum luminosities weaker by 0.4 to more than 0.8\,dex (2$\sigma$ limit) given the observed CO luminosity. The dashed line and grey shaded regions show the best-fit relation and inner $\pm$1,2$\sigma$ scatter from \citet{scoville23}.
}\label{fig:observables}
\end{figure}

Before considering any physical quantities, we first ask whether the CO or dust properties of our $z\sim1$ quiescent galaxies are empirically unusual. Figure~\ref{fig:observables} compares the observed CO and rest-frame 850\,\um luminosities to the sample of star-forming galaxies assembled by \citet{scoville23}. As in that work, we assume $\Tdust = 25$\,K and a long-wavelength dust spectral index $\beta = 2$ to k-correct our rest-frame $\approx$500\,\um measurements to rest-frame 850\,\um. We verified that our assumptions to calculate gas and dust masses from the luminosities recover $\gdr \approx 100$ when using the \citet{scoville23} observables. The comparison sample contains a wide diversity of star-forming galaxies, from typical $z\sim0$ spirals to local and high-redshift IR-luminous sources. 

It is obvious from Fig.~\ref{fig:observables} that our quiescent galaxies are clear outliers, with dust luminosity lower by a factor of at least 0.4\,dex to more than 0.8\,dex given their CO luminosity.\footnote{The few literature galaxies we can place on this plot are consistent with this result; see also Fig.~\ref{fig:gdrssfr}.} We note that the comparison sample uses exclusively CO(1--0) measurements while we use CO(2--1), but the deficit in dust emission is larger than any plausible change due to CO excitation (maximum $\approx$25\%; Section~\ref{caveats}). The deficit of dust luminosity is well outside the scatter of the CO-dust relation derived by \citet{scoville23}, which we find to be $^{+0.17}_{-0.20}$\,dex (inner 68th percentile) and $^{+0.28}_{-0.32}$\,dex (inner 95th percentile). From a purely observational standpoint, then, we conclude that our sample of $z\sim1$ quiescent galaxies show dust emission significantly weaker than expected. We subsequently interpret this difference as arising from abnormally large gas-to-dust mass ratios in the quiescent galaxies.

\subsection{Gas and Dust Fractions}

%%%%%%%%%%%%%%%%%%%%%%%%%%%%%%%%%%%%%%%%%%%%%%%%%%%%%%%%%%%%%%%%%%%%%%%%%%%%%%%%%
%%% FIGURE - sSFR vs fH2 and fdust
\begin{figure*}[t]
\begin{centering}
\includegraphics[width=\textwidth]{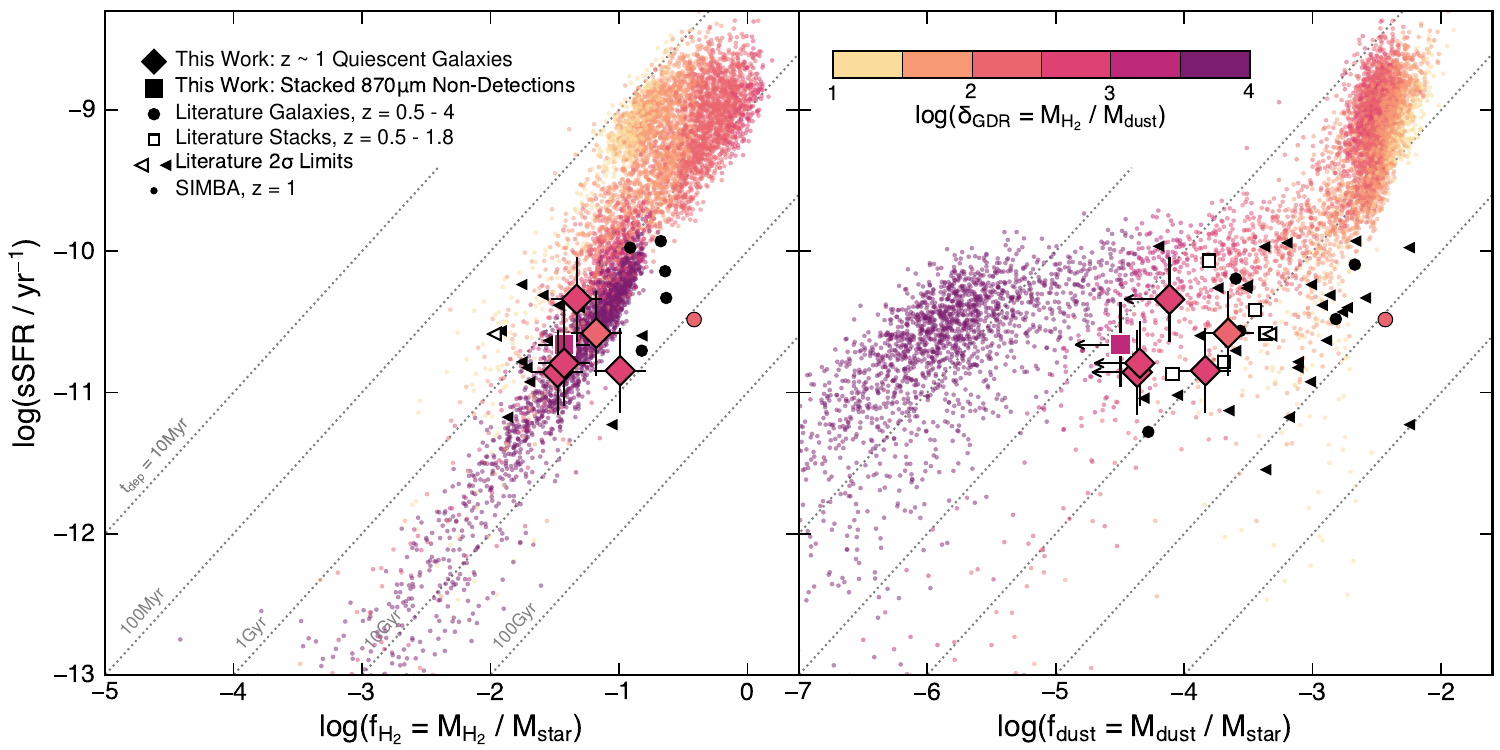}
\end{centering}
\caption{
SIMBA exhibits a tight relationship between sSFR and \fht but often a sharp decrease in \fdust in quenched galaxies \citepalias{whitaker21b}. These trends are in broad agreement with observations, including our own.
Our measurements are shown as large diamonds, color-coded according to \gdr, and the stacked non-detection as a large square.
Individual quiescent galaxies from the literature are shown as circles (for detections) or left-facing triangles (2$\sigma$ upper limits). Literature stacking experiments are shown similarly as open symbols. The galaxy ALMA.14 \citep{hayashi18} is the only literature source with a measured \gdr.
Galaxies from SIMBA are shown as small dots, color-coded according to \gdr.
The dotted lines in the left-hand panel are lines of constant gas depletion time $\tdep = \MHt$/SFR; the lines in the right-hand panel show the same if one also assumes $\gdr=100$.
Note that the \gdr color scaling uses 2$\sigma$ limits where needed; \gdr is more directly visualized in Figure~\ref{fig:gdrssfr}.
}\label{fig:fhtfdust}
\end{figure*}

The relationship between sSFR, $\fht = \MHt/\Mstar$, and $\fdust = \Mdust/\Mstar$ is shown in Figure~\ref{fig:fhtfdust}, for both observed high-redshift quiescent galaxies and the SIMBA simulated galaxies. The normalization by \Mstar is intended to lessen mass-related variations and uncertainties in the gravitational lens models for the lensed objects in the literature sample. The comparison to SIMBA would not change by using either the $z=0.5$ or $z=2$ simulation outputs (above this redshift SIMBA produces few quiescent galaxies in our mass range). 

As previously noted by \citetalias{whitaker21b}, there is a tight relationship between sSFR and \fht but enormous scatter between sSFR and \fdust. For the SIMBA galaxies, the trend with \fht is easily explained by the Schmidt-like criterion for star formation used in the simulation, which results in a power-law relationship between SFR and \MHt with an index of 1.5.\footnote{This is why the redshift of the simulation snapshot used does not greatly affect our discussion in this section.} The large scatter with \fdust, on the other hand, is a result of the prescriptions for dust grain production, destruction, and grain growth used in SIMBA. \citetalias{whitaker21b} show that the largest driver of the difference in \gdr between star-forming and quiescent galaxies is grain growth in the ISM of star-forming galaxies, which keeps \gdr low. In other words, dust destruction affects star-forming and quiescent galaxies alike, but only star-forming galaxies efficiently counter this destruction. In lower-mass quiescent galaxies, prolonged ISM grain growth can eventually rebuild a modest dust mass \citep{lorenzon25a}, but at least in SIMBA this is exceedingly rare at the masses we consider ($\log(\Mstar/\Msun)>10$).

The gas properties of observed high-redshift quiescent galaxies are generally compatible with the trends seen in SIMBA. The gas fractions of the observed galaxies appear to have somewhat higher scatter than the simulated ones -- when upper limits are considered, we see outliers to both higher and lower \fht than typical in SIMBA. Some of this scatter, no doubt, is due to the challenges in measuring SFRs at such low values and the uncertainty in the gas mass conversions we adopted. A similar result was found for the EAGLE simulation \citepalias{spilker18b}. The most gas-rich galaxy here is ALMA.14, a member of a $z=1.46$ cluster with unusually high gas and dust fractions \citep{hayashi18} (all other detected cluster members are likely star-forming).

The right panel of Fig.~\ref{fig:fhtfdust} also shows reasonable agreement between the observed dust mass fractions and those in SIMBA. This would remain true if the observed SFRs have been overestimated by up to a factor of a few, because there is very large scatter in \fdust at a given sSFR. The observational constraints on the dust fraction in individual objects are often more limited than for the gas fraction. This is simply because the literature measurements are a combination of studies that specifically observed the dust continuum at relatively high rest-frame frequencies, and those that were primarily aimed at detecting CO or \ci line emission at lower rest-frame frequencies. For these latter sources, the dust estimates are naturally less constraining due to the steep slope of the dust SED; these programs were neither designed nor intended to place strong constraints on \fdust. For our five target galaxies, both the detected and undetected sources are consistent with the expectations from SIMBA.

This exercise is primarily intended to illustrate that the landscape of current observations is broadly compatible with the expectations of modern cosmological simulations (at least one that explicitly models dust production, growth, and destruction).

\subsection{Unusually High Gas-to-Dust Ratios Observed in Quiescent Galaxies} \label{highgdr}

%%%%%%%%%%%%%%%%%%%%%%%%%%%%%%%%%%%%%%%%%%%%%%%%%%%%%%%%%%%%%%%%%%%%%%%%%%%%%%%%%
%%% FIGURE - sSFR vs GDR
\begin{figure}[t]
\begin{centering}
\includegraphics[width=\columnwidth]{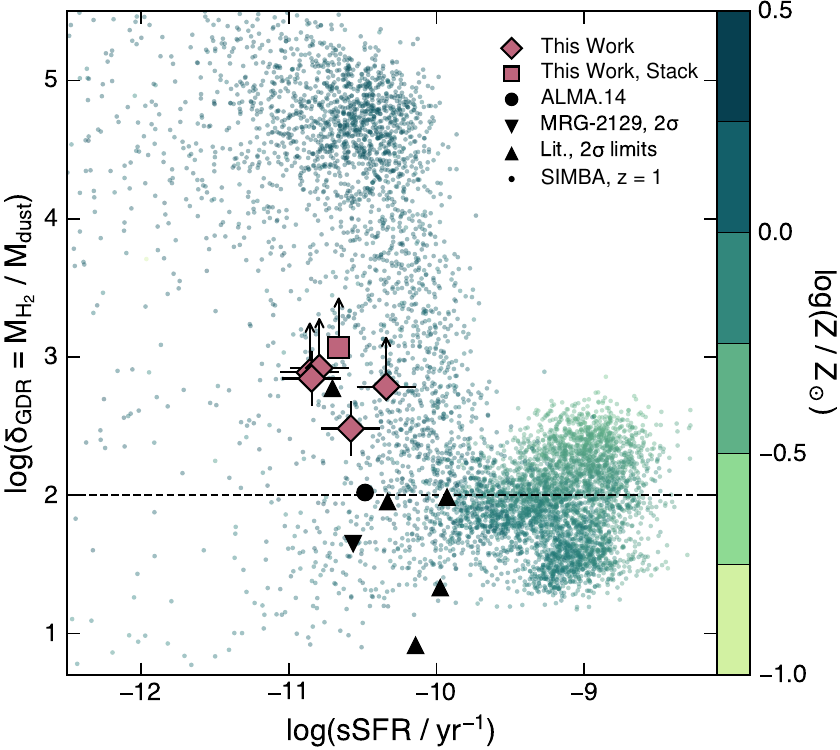}
\end{centering}
\caption{
We find values of \gdr much higher than the typically-assumed value of 100 in our sample of $z\sim1$ quiescent galaxies -- even the most `dust-rich' object has $\gdr \approx 300$, and we find a 2$\sigma$ limit $\gdr \gtrsim 1200$ from our stack of undetected galaxies. Our observations, and others from the literature, are generally consistent with predictions from SIMBA. We highlight ALMA.14 \citep{hayashi18} as the only other object with a measured \gdr, and MRG-2129 \citep{whitaker21a,morishita22} with a 2$\sigma$ \textit{upper} limit $\gdr \lesssim 50$.
Symbols otherwise are as in Fig.~\ref{fig:fhtfdust}; all limits are 2$\sigma$.
}\label{fig:gdrssfr}
\end{figure}

Our observations were specifically designed to measure the molecular gas-to-dust ratio \gdr in high-redshift quiescent galaxies, which we show in Figure~\ref{fig:gdrssfr}. While our targets have modest gas fractions $\fht \approx$3--10\% (somewhat by construction, since we required CO detections for higher-frequency continuum followup), they are unexpectedly dust-poor relative to standard empirical calibrations \citep[e.g.][]{scoville23}. Under our fiducial assumptions, the two galaxies with detected dust continuum, C130284 and C132776, have $\gdr = 300 \pm 70$ and $700 \pm 250$, respectively.\footnote{Incidentally, C130284 appears to be a spiral quiescent galaxy with a nearby star-forming companion visible in HST and JWST imaging. This may suggest that it quenched slowly \citep{park22} and has maintained a dust reservoir through long-term grain growth \citep{lorenzon25a}.} For the remaining sources, we place $2\sigma$ lower limits of $\gdr > 610-830$, and $>$1200 from the stack of undetected galaxies. These estimates are substantially larger than the typical $\gdr \sim 100-200$  assumed in the literature, but are a direct consequence of the unexpectedly weak continuum emission we observe.

There are far fewer literature comparison galaxies in Fig.~\ref{fig:gdrssfr} because this plot requires a detection in either a gas tracer and/or dust. The only other quiescent galaxy detected in both gas and dust is ALMA.14 \citep{hayashi18}; under our assumptions, we find $\gdr \approx 100$ for this source. Most remaining galaxies have not-very-constraining lower limits -- these are objects detected in a low-frequency CO transition, where dust constraints were not the primary goal of the original observations. The only galaxy with a \gdr upper limit is MRG-2129 at $z=2.14$, which has been detected at multiple ALMA frequencies but remained undetected in \ci($^3$P$_2$ -- $^3$P$_1$) emission \citep{whitaker21a,morishita22}. Interestingly, this galaxy is the only one from the parent \citet{newman18a} sample of lensed quiescent galaxies with evidence for an AGN from its optical nebular lines; whether this is causal or coincidental is not clear.

Our observation of `exotic' high values of \gdr in massive, presumably metal-rich quiescent galaxies confirms a prediction of SIMBA, as pointed out by \citetalias{whitaker21b}. Below a critical value of $\log(\mathrm{sSFR}/\mathrm{yr}^{-1}) \approx -10$, gas and dust are essentially uncorrelated in SIMBA, resulting in values of \gdr that span several orders of magnitude to reach $>10^5$. This effect occurs in massive, metal-rich galaxies; the increase is not due to the same physics responsible for the metallicity dependence of \gdr observed in metal-poor galaxies. Observational studies like ours cannot hope to reach such deep limits in reasonable time, but our measurements and those from the literature are consistent with a scenario in which \gdr varies wildly in low-SFR galaxies, from $\lesssim$50 (observed in MRG-2129) to at least $\gtrsim$1200 (from our stack of undetected galaxies). A similar result was recently found by \citet{lorenzon25b} in a sample of quiescent galaxies at $z\sim0.3$, in which \gdr reached $\approx$800 and spanned a factor of $\approx$16$\times$.

\subsection{Are Quiescent Galaxies Gas-Rich or Dust-Poor?} \label{trendmstar}

%%%%%%%%%%%%%%%%%%%%%%%%%%%%%%%%%%%%%%%%%%%%%%%%%%%%%%%%%%%%%%%%%%%%%%%%%%%%%%%%%
%%% FIGURE - fH2 and fdust vs Mstar
\begin{figure*}[t]
\begin{centering}
\includegraphics[width=\textwidth]{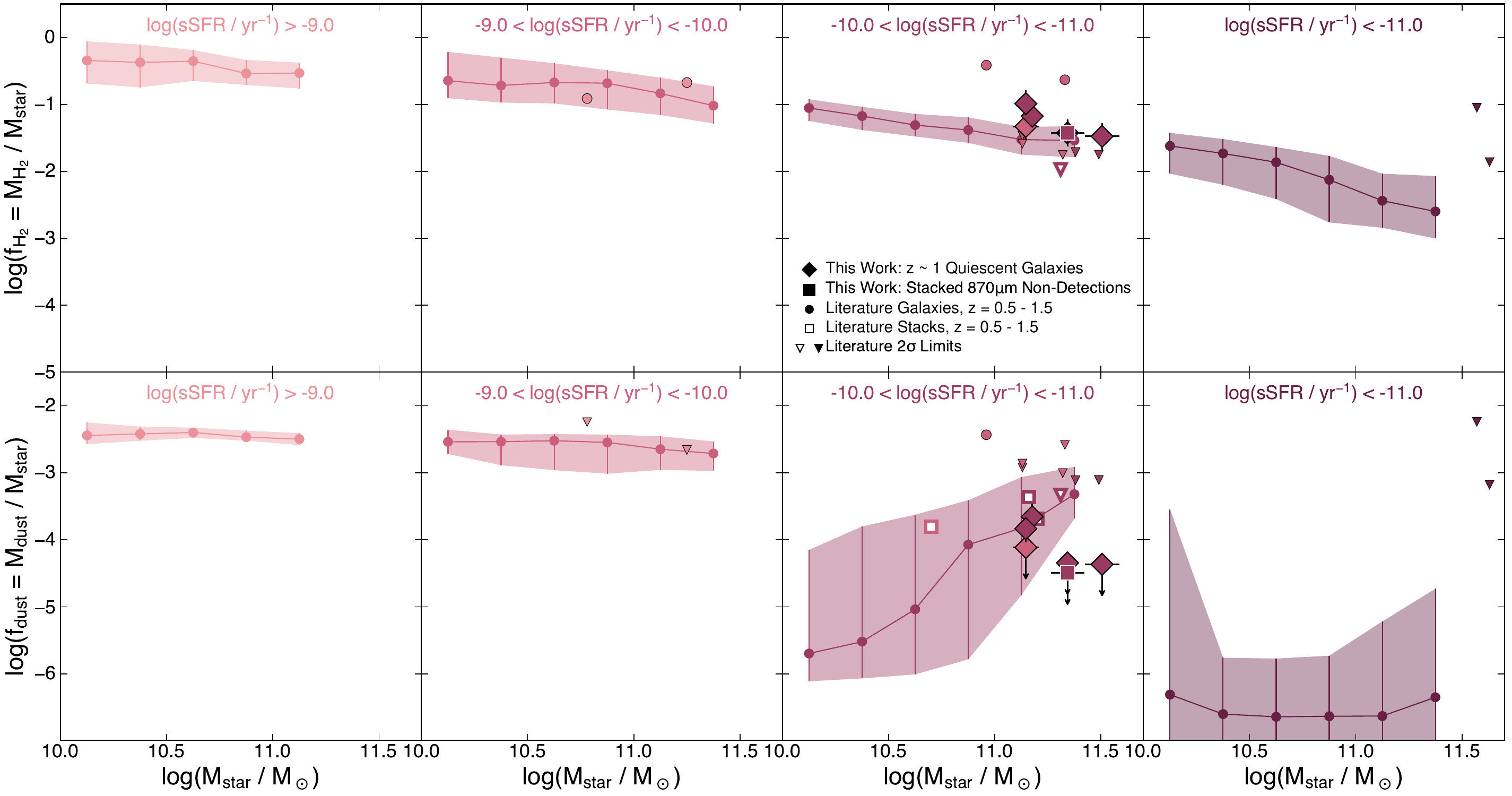}
\end{centering}
\caption{
In contrast to the mild decline of \fht with increasing \Mstar at all sSFRs, SIMBA predicts no clear trend between \fdust and \Mstar for quiescent galaxies. Observed galaxies scatter both above and below the expected \fht, while the most massive quiescent galaxies have lower \fdust than predicted.
Circles and errorbars / shaded regions correspond to the median and 1$\sigma$ interval of SIMBA simulated galaxies, while diamonds, squares, and small circles are observed galaxies as in previous figures. All upper limits (downward triangles) are 2$\sigma$. 
}\label{fig:mstar}
\end{figure*}

Although our sample galaxies were selected to have CO detections, they are not necessarily unusual in their gas properties. While observational constraints are limited, the gas contents are consistent with extrapolations of scaling relations developed for star-forming systems \citepalias{spilker18b}. On the other hand, the H$_2$ masses are higher than expected by the EAGLE simulation \citepalias{spilker18b}, and possibly also SIMBA if our SFRs are overestimated (Figure~\ref{fig:fhtfdust}).

The picture for the dust contents is also unclear. Studies of individual objects often find low dust fractions \citep{whitaker21a}, while some stacks of the dust continuum in large samples find higher fractions \citep[e.g.][]{gobat18,magdis21,blanquezsese23}. Our results only exacerbate these differences: our targets have relatively high gas masses, traced by CO(2--1), but weak continuum emission and consequently low dust masses. 

One possible resolution, suggested by \citet{blanquezsese23}, is that the stacking experiments typically probe lower stellar masses than most individual targets. If the dust (and gas) fraction is mass-dependent, as observed in typical star-forming galaxies \citep[e.g.][]{liu19a}, this selection effect could explain the discrepancy because lower-mass galaxies have more dust than at higher masses.

Figure~\ref{fig:mstar} shows the gas and dust fractions from SIMBA and observed quiescent galaxies as a function of \Mstar, now restricting the literature comparison sample to galaxies at $0.5 < z < 1.5$ to minimize redshift evolution. SIMBA shows a mild decline in molecular gas fraction with stellar mass for galaxies above, near, and below the star-forming sequence, in agreement with scaling relations derived for massive star-forming galaxies across redshifts \citep[e.g.][]{liu19a}. The dust fraction follows a similar behavior for galaxies near the star-forming sequence, but the efficient destruction of dust as galaxies become quiescent breaks the correspondence between gas and dust. In SIMBA, the dust fraction no longer smoothly declines with increasing stellar mass for quiescent galaxies.

Observed galaxies show broad agreement with the trends in SIMBA, with some nuance. We first notice that all literature stacks of dust fraction are within the scatter of the SIMBA trends; these galaxies are not actually more dust-rich than expected. Second, from the individual galaxies we see deviations toward both high \fht and low \fdust compared to SIMBA. The literature dust stacks would also have mildly higher-than-expected \fht when assuming $\gdr=100$ and moreso for values $>$100, in agreement with the CO-based \fht measurements. Some objects (with only upper limits on \fht), in contrast, appear to have less molecular gas than expected. The deviations of molecular gas in both directions are, as noted in Fig.~\ref{fig:fhtfdust}, evidence that there is more scatter in the cold gas properties of real quiescent galaxies than in the simulation.

It is interesting to consider the implications if the quiescent galaxy SFRs have been overestimated -- a likely possibility, given the challenges in measuring low-level star formation. The observed points in Fig.~\ref{fig:mstar} would then move to the rightmost panel. Galaxies already detected in gas and dust would uniformly fall above the SIMBA expectations. Those galaxies with limits on \fdust that presently fall below the SIMBA expectation would become more consistent with the simulation, but at the cost of a worsened excess of cold gas compared to the expectation from SIMBA.

\subsection{Impact of Assumptions on Gas and Dust Masses} \label{caveats}

We found unexpectedly high gas-to-dust ratios $\gdr \gtrsim 300-1200$ in massive quiescent galaxies at $z\sim1$, but we now consider whether any of our fiducial assumptions in the gas and/or dust mass calculations could be systematically biased enough to recover $\gdr \sim 100$. We applied the same set of assumptions uniformly to all literature observations of quiescent galaxies, and we recall that our sample empirically has unusually weak dust emission (Fig.~\ref{fig:observables}). For the quiescent galaxies to have more `normal' \gdr, then, we must invoke a scenario in which the conversions from observable to physical quantities are different for quiescent and star-forming galaxies.

Variations in the dust temperature \Tdust and CO excitation factor $r_{21}$ are the easiest to rule out, because neither can produce the magnitude of the difference we observe. Increasing $r_{21}$ from 0.8 to 1.0 (i.e. thermalized CO emission) would only decrease the CO(1--0) luminosity by 25\%, far lower than the factors of 2.5 to $>$6 needed to bring the quiescent galaxies into agreement. Similarly, substantial literature argues that, in contrast to the light-weighted \Tdust, the mass-weighted \Tdust has little variation \citep[e.g][and references therein]{scoville16}. Lowering \Tdust from 25 to 20\,K \citep[e.g.][]{magdis21} only increases \Mdust by a factor of $\approx$40\%, which is likewise too small to bring \gdr down to 100-200. In part, this shift is small because our sample galaxies lie at relatively low redshift, and our ALMA observations probe a rest-frame wavelength firmly on the Rayleigh-Jeans tail of the dust emission \citep{liang19,cochrane22}.

Another alternative that we consider unlikely is that the CO emission arises solely from the molecular gas in the ISM, but dust emission traces both H$_2$ and neutral atomic HI. Accounting for HI (in both star-forming and quiescent galaxies) would increase the (atomic plus molecular) gas-to-dust ratio overall. To get the total gas-to-dust ratio to match between populations would require that the fraction of gas in the molecular phase be much higher in quiescent galaxies than star-forming galaxies, contrary to expectations (indeed it is common to neglect HI entirely for high-redshift star-forming galaxies; see \citealt[][for a discussion]{tacconi20}). We verify that, at least in SIMBA, galaxies with abnormally high molecular \gdr also have abnormally high total (HI$+$H$_2$) \gdr, implying that it is the dust that has changed and not the balance of ISM phases.

The CO--H$_2$ conversion factor is a much larger hammer at our disposal. We assumed a `Milky Way like' $\alphaco = 4.4$, typical of galaxies in which the molecular gas is confined to a collection of virialized clouds \citep{bolatto13}. Substantially lower values, $\alphaco \sim 1$, are inferred in local IR-luminous galaxies. These sources are almost ubiquitously gas-rich mergers, and the decrease in \alphaco is primarily due to warmer gas and decreased CO line opacity that allows more luminosity to escape for a given molecular mass. In quiescent galaxies at $z\sim0$, \citet{crocker12} find line ratios using the optically-thin $^{13}$CO isotopologue and dense gas tracers HCN and HCO$^+$ that are consistent with local spirals, implying that the prevailing ISM conditions in local quiescent galaxies are not like those in nearby mergers. Lowering \alphaco by a factor of 4.4 would bring most of our measurements into the typical range of \gdr. While we cannot rule out this scenario, it seems unlikely that our low-SFR galaxies predominantly have molecular gas in the warm, excited, low-opacity phase typical of local IR-luminous mergers.

Lastly, if we cannot change the gas, we must change the dust -- specifically, the dust absorption coefficient $\kappa_\nu$ would have to be lower in quiescent than star-forming galaxies so that a given continuum luminosity corresponds to a higher dust mass. We explored a number of different dust emissivity parameterizations from the literature to understand its impact \citep{hildebrand83,james02,dunne03,draine07,guillet18,hensley23}. Of these, only the emissivity of \citet{draine07} increased the dust masses, by $\approx$50\% compared to our fiducial assumption, again smaller than the deficit of dust emission we observe.

To summarize, while there are significant uncertainties in the computation of both molecular gas and dust masses, we have no substantive reason to alter our fiducial assumptions. Empirically, the dust emission is unusually faint in our target galaxies. We would need \textit{all} of the above caveats to be true to bring \gdr down to the standard value -- high-excitation CO, cold and less emissive dust, and a lower \alphaco. Unless there truly are order-of-magnitude systematic differences in the CO and/or dust properties of quiescent vs. star-forming galaxies, the deficit of dust emission we observe is a robust indication that \gdr is higher in the former population at $z\sim1$.

\section{Discussion and Conclusions} \label{conclude}

We have presented the first systematic study of the gas-to-dust ratio \gdr in quiescent galaxies outside the local universe, finding that these sources have weaker dust emission than expected given the CO luminosity by a factor of 2.5 to more than 6 (Fig.~\ref{fig:observables}). We interpret this as evidence that \gdr is substantially higher than typically assumed in the literature by similar factors (Fig.~\ref{fig:gdrssfr}), casting doubt on prior dust-based estimates of the cold gas contents of quiescent galaxies. Even though these galaxies are massive and therefore presumably metal-rich, we find evidence that \gdr can reach at least 1200. Our result is in agreement with recent work by \citet{lorenzon25b}, who found values up to $\gdr\approx800$ in a sample of lower-redshift quiescent galaxies using similar CO and dust continuum observations.

It is not clear whether our target galaxies, preselected to have CO detections, are typical in their gas and/or dust properties. We show that other observations of the molecular gas and dust in high-redshift quiescent galaxies are generally consistent with our observations, and with the SIMBA cosmological simulation. While SIMBA recovers a smooth, mildly declining gas fraction with increasing stellar mass for all sSFRs, the dust fraction does not show the same behavior (Fig.~\ref{fig:mstar}). We show that both prior dust stacking results and targeted observations of individual galaxies are consistent with SIMBA, especially if the SFRs of these quiescent galaxies are overestimated. 

While acknowledging our small sample size, we are faced with the sobering conclusion that observations of the dust continuum cannot reliably be used to trace the cold gas contents of high-redshift quiescent galaxies. Regardless of conversion factors, these galaxies have unusually weak dust emission. This is supremely inconvenient, because continuum observations can reach much deeper limits in gas mass for a given observing time than low-order CO transitions, if typical assumptions about \gdr hold. That advantage disappears if dust is an order of magnitude (or more) less abundant than in star-forming galaxies. 

We argue that a much fuller understanding of the variations in \gdr in high-redshift quiescent galaxies is needed, in particular whether there are any observable properties that correlate with \gdr (stellar age, perhaps). Given the relative expense, in our view the most obvious path forward is that any and all CO-based studies of quiescent galaxies be followed up in (cheaper) continuum observations. Additional CO line observations of dust-detected or undetected galaxies will also prove valuable to estimate the full range of variations in \gdr. It is our hope that \gdr can be understood well enough in the future that continuum observing campaigns can be considered robust probes of the cold gas in low-SFR galaxies. Unfortunately, we do not yet have sufficient understanding to convert observed submillimeter fluxes into gas masses in distant quiescent galaxies.

\begin{acknowledgements}

JSS thanks Sirio Belli for providing the NOEMA data from his 2021 work. JSS, VRD, and KAS gratefully acknowledge support from NSF-AAG\#2407954 and 2407955, and NRAO SOSPA11-006. KEW and SB gratefully acknowledge support from NSF-CAREER\#2144314. MEV acknowledges support from from NSF grant AST 22-06164 and the Center for Astrophysical Surveys Graduate Fellowship. PFW acknowledges funding through the National Science and Technology Council grant 113-2112-M-002-027-MY2. 
This paper makes use of the following ALMA data: 2016.1.00790.S, 2018.1.01739.S, 2022.1.00642.S. ALMA is a partnership of ESO (representing its member states), NSF (USA) and NINS (Japan), together with NRC (Canada), MOST and ASIAA (Taiwan), and KASI (Republic of Korea), in cooperation with the Republic of Chile. The Joint ALMA Observatory is operated by ESO, AUI/NRAO and NAOJ. The National Radio Astronomy Observatory is a facility of the National Science Foundation operated under cooperative agreement by Associated Universities, Inc.
This research has made use of NASA's Astrophysics Data System.

\end{acknowledgements}

\facility{ALMA}

\software{
CASA \citep{casa22},
CARTA \citep{carta24},
\texttt{astropy} \citep{astropy18},
\texttt{matplotlib} \citep{hunter07}}

%\clearpage
\bibliographystyle{aasjournalv7}
%\bibliography{/Users/jspilker/Research/master.bib}

\end{CJK*}
\end{document}